\documentclass{article}
\usepackage{graphicx}  
\usepackage{amsmath}   
\usepackage[compress]{cite}
\usepackage{amssymb}   
\usepackage{bm} 
\usepackage{dcolumn}
\usepackage{color}
\usepackage{mathrsfs}
\usepackage{amsfonts}
\usepackage{varioref}
\usepackage{textcomp}

\RequirePackage[colorlinks,citecolor=blue,urlcolor=magenta,linkcolor=blue]{hyperref}
\allowdisplaybreaks

\labelformat{section}{Section #1} 
\labelformat{subsection}{Section #1} 
\labelformat{subsubsection}{Section #1}
\labelformat{subsubsubsection}{Section #1}
\labelformat{equation}{Eq.~(#1)} 
\labelformat{figure}{Fig.~#1} 
\labelformat{subfigure}{Fig.~\thefigure#1} 
\labelformat{table}{Table~#1} 
\labelformat{appendix}{Appendix #1}
\newcounter{mnotecount}[section]

\renewcommand{\themnotecount}{\thesection.\arabic{mnotecount}}

\newcommand{\mnotex}[1]
{\protect{\stepcounter{mnotecount}}$^{\mbox{\footnotesize
$
\bullet$\themnotecount}}$ \marginpar{
\raggedright\small\em
$\!\!\!\!\!\!\,\bullet$\themnotecount: #1} }

\begin{document}
\title{Hawking radiation through tunneling from a hot NUT-Kerr-Newman-Kasuya-Anti-de Sitter black hole}
\author{ Heisnam Shanjit Singh \footnote{heisnam.singh@rgu.ac.in}$~^{1}$
Chiranjeeb Singha \footnote{chiranjeeb.singha@iucaa.in}$~^{2}$, and Sraban Kumar Upadhyaya\footnote{sraban.upadhyaya@rgu.ac.in}$~^{1}$\\$^{1}$ \small{Department of Physics, Rajiv Gandhi University, Rono Hills, Doimukh Papumpare 791112, Arunachal Pradesh India}\\
$^{2}$\small{Inter-University Centre for Astronomy and Astrophysics, PostBag 4, Ganeshkhind, Pune 411007, Maharashtra, India}}
\date{\today}
\maketitle

\begin{abstract}
In this research work, we examine the tunneling phenomenon of a charged Dirac particle emerging from the thermal horizon of a hot NUT-Kerr-Newman-Kasuya-Anti-de Sitter (HNKNK-AdS) black hole. Using the tunneling formalism, we derive the Hawking temperature associated with the charged Dirac particle at the horizon of the HNKNK-AdS black hole. Our findings reveal that the effective Hawking temperature depends on various black hole parameters, including electric and magnetic charges, magnetic mass, the cosmological constant, and angular momentum. Additionally, the Hawking radiation for the Kerr-Newman black hole diminishes slightly, and a minor correction to the thermal spectrum is identified. We also discuss the heat capacities of the HNKNK-AdS and find that there exists an infinite number of discontinuity implying that the black hole system becomes unstable as the smaller the event horizon of the black hole and hence deduces the well known heat capacity for the Kerr-Newman black hole.
\end{abstract}

\section{Introduction}\label{sec1}
Employing quantum field theory in curved spacetime, Hawking \cite{Hawking1975} identified the emission of thermal radiation, now known as Hawking radiation, through the event horizon of a black hole. Building on this, Gibbons and Hawking \cite{Gibb} extended the analysis to cosmological event horizons, demonstrating thermal radiation's emergence due to quantum effects in spacetime. This foundational discovery bridges general relativity, quantum field theory, and statistical thermodynamics, sparking interest in Hawking radiation as a potential window into quantum gravity.

Subsequent studies introduced key methods, such as the tunneling method \cite{Parikh, Kraus}, the Hamilton-Jacobi method \cite{Angheben} extending Padmanabhan's complex path analysis \cite{Padhmanabhan},  and the gravitational anomaly method \cite{Christensen}, to explore the dynamical origins of Hawking radiation. For instance, Zhang and Zhao applied the tunneling method to Reissner-Nordström and Kerr-Newman black holes \cite{zhang, zheng}. An improved Hawking temperature of a Dyonic black hole using a bosonic tunnelling strategy based on the Hamilton-Jacobi technique was investigated using the semi-classical method, the Wentzel-Kramers-Brillouin (WKB) approximation, and the Lagrangian equation in the presence of quantum gravity as observed in the generalised uncertainty principle (GUP)\cite{AliRi}. Incorporating the GUP-induced changes to the emitted particle's behaviour into the Hamilton-Jacobi equation, the modified Hawking temperature of the Schwarzschild black hole inside the bumblebee gravity model was discussed, calculating the tunnelling probability \cite{Sakallı}. In Starobinsky–Bel–Robinson (SBR) gravity, Annamalai, D. and Pandey, A. \cite{Annamalai} examined the Hawking radiation of a Schwarszchild-type black hole and the corrected Hawking temperature using the tunnelling method. Siahaan, Haryanto M., et al.\cite{Siahaan} calculated Hawking radiation using the tunneling picture for the magnetised Taub–NUT spacetime with the Manko–Ruiz parameter. Hawking temperature and modified fermion tunnelling radiation of the stationary axisymmetric Kerr-TAUB-NUT black hole in the curved space-time were calculated by applying the correction of the Lorentz dispersion relation \cite{Liu}. Murata and Soda \cite{Keiju} examined Hawking radiation from rotating black holes through gravitational anomalies. The phenomenon arises from spontaneous particle-antiparticle pair creation near the event horizon, with the outer horizon radiating similarly to a Schwarzschild black hole. In contrast, the radiation dynamics of inner horizons remain less understood. Using the Klein–Gordon field equation, Wu and Cai \cite{wu2000, wu2004, wu2005} investigated inner horizon radiation, reporting negative temperatures, a controversial finding in black hole thermodynamics \cite{paddy1, paddy2}. Conversely, Peltola and Mäkelä \cite{peltola} argued for positive effective temperatures in maximally extended Reissner-Nordström spacetimes. The role of inner horizons in black hole thermodynamics remains unresolved.
Black hole entropy presents a similarly complex challenge \cite{Frolov, Wald}. According to Bekenstein \cite{Bekenstein}, entropy is proportional to the event horizon's surface area, while temperature is related to surface gravity \cite{Bardeen}. The Nernst theorem posits that entropy should vanish as temperature approaches zero \cite{Wald}. However, entropy does not vanish at absolute zero for black holes with multiple horizons, such as Kerr black holes, unless the system is treated as two subsystems (outer and inner horizons), satisfying the theorem when their contributions are included \cite{He}. Recent studies by Ren \cite{Ren, RenJun} examined Hawking radiation contributions from inner horizons, and global temperatures for multi-horizon spacetimes have been proposed \cite{Volovik:2021iim, Volovik:2021upi, Singha:2021dxe, Choudhury:2004ph}.

In multi-horizon spacetimes, thermodynamic behaviour diverges from single-horizon cases. Correlations between horizons in Reissner-Nordström and Kerr black holes have been suggested to alter temperature and entropy, deviating from traditional outer-horizon-only models \cite{Volovik:2021upi, Volovik:2021iim}. These correlations challenge simplistic summations of horizon entropies. Chiranjeeb's work \cite{Singha:2021dxe} further examined global temperature and entropy for Schwarzschild-de Sitter, Reissner-Nordström-de Sitter, and rotating BTZ spacetimes, suggesting a systematic approach to derive global thermodynamic quantities based on horizon contributions. Additional studies have investigated Hawking radiation and tunneling for charged particles in complex spacetimes, such as Reissner-Nordström-Taub-NUT and Kerr-Newman black holes with electric and magnetic charges \cite{Ali, Juan}. Subsequently, several works on Hawking radiation have been studied in recent literature. For instance, Media et al. \cite{Media:2024tyg} examined how the Lorentz symmetry violation affected the Hawking temperature and heat capacity of the KN-AdS black hole surrounded by perfect fluid.

This paper investigates Hawking radiation and heat capacity for Dirac-charged particles emitted from the hot NUT-Kerr-Newman-Kasuya-Anti-de Sitter (HNKNK-AdS) black hole via the tunneling process. This spacetime, defined by parameters including mass, angular momentum per unit mass, cosmological constant, and electric and magnetic charges, is particularly intriguing due to the theoretical presence of magnetic monopoles. The investigation of quantum effects in the hot NUT-Kerr-Newman-Kasuya-Anti-de Sitter (HNKNK-AdS) black hole is interesting because of the fact that magnetic monopoles exist has been given on the grounds of the symmetry that they would introduce in the field equations of electromagnetism. We compute the tunneling of the Dirac particle to determine whether Hawking's temperature depends on the contributions of the event horizon of the black hole and, hence, possible deduction of the Hawking temperature in the case of the Kerr-Newman black hole.

The paper is organized as follows: In section 2, we review the HNKNK-AdS black hole in curved spacetime. In section 3, we derive Hawking radiation using the Dirac equation and investigate global temperature for this spacetime. We also discuss the heat capacities of the HNKNK-AdS and Kerr-Newman black holes in the same section. In section 4, we discuss our findings. Throughout, we use natural units ($G = c = \hbar = 1$) unless specified otherwise. 

\section{Hot NUT-Kerr-Newman-Kasuya-Anti-de Sitter space-time}\label{sec2}

We study an HNKNK-AdS black hole characterized by the parameters: mass $ M$, specific angular momentum $ a = \frac{J}{M}$, NUT (magnetic mass) parameter $ n$, electric charge $ Q_e$, magnetic monopole parameter $ Q_m$, and the cosmological constant $ \Lambda$. According to an exact solution for a rotating dyon black hole, the metric for the four-dimensional HNKNK-AdS black hole in Boyer-Lindquist coordinates is expressed as \cite{kasuya}:

\begin{equation}
ds^2 = g_{tt} dt^2 + g_{rr} dr^2 + g_{\theta \theta} d\theta^2 + g_{\phi \phi} d\phi^2 + 2g_{t\phi} d\phi dt,
\end{equation}
where the metric components are given by:
\begin{equation}
\begin{aligned}
g_{tt} &= -\frac{\Delta - a^2 \Delta_\theta \sin^2\theta}{\Sigma},\\  \quad 
g_{t\phi}& = g_{\phi t} = -\frac{h \Delta - a \xi \Delta_\theta \sin^2\theta}{\Xi \Sigma},\\ \quad 
g_{\theta\theta}& = \frac{\Sigma}{\Delta_\theta}, \\
g_{rr} &= \frac{\Sigma}{\Delta},\\ \quad 
g_{\phi\phi} &= -\frac{h^2 \Delta - \xi^2 \Delta_\theta \sin^2\theta}{\Xi^2 \Sigma}.
\end{aligned}
\end{equation}
The auxiliary functions are defined as:
\begin{equation}
\begin{aligned}
\Xi &= 1 - \frac{a^2}{y^2},\\ \quad 
h & = a\sin^2\theta - 2n\cos\theta,\\ \quad 
\Sigma &= r^2 + (n + a\cos\theta)^2,\\ \quad \xi^2 &= a^2 + r^2 + n^2, \\
\Delta &= \xi^2 \left( \frac{5n^2 + r^2}{y^2} + 1 \right) - 2(Mr + n^2) + Q^2,\\ \quad 
\Delta_\theta &= 1 - \frac{a^2 \cos^2\theta}{y^2}.
\end{aligned} 
\end{equation}
Here, the negative cosmological constant is related to $ y$ by $ y^2 = -3/\Lambda$, and the total charge is $ Q^2 = Q_e^2 + Q_m^2$. The electromagnetic field tensor for this metric is:
\begin{equation}
F = \frac{Q}{\Sigma^2 \Xi} \left[ r^2 - (n + a\cos\theta)^2 \right] dr \wedge (dt - h d\phi) - \frac{2Qr}{\Sigma^2 \Xi} (n + a\cos\theta) \sin\theta \, d\theta \wedge (-a dt + \xi^2 d\phi).
\end{equation}
The non-vanishing components of the electromagnetic potential are:
\begin{equation}
A_\mu = -\frac{Qr}{\Xi \Sigma} \delta_{\mu t} + \frac{Qr h}{\Xi \Sigma} \delta_{\mu \phi}.
\end{equation}
The event horizons of the black hole are determined by the roots of $ \Delta = 0$. Near the event horizon $ r_h$, $ \Delta$ can be expanded as:
\begin{equation}
\Delta = (r - r_h) \frac{d\Delta}{dr}\Big|_{r=r_h} = (r - r_h) \Delta'\Big|_{r=r_h},
\end{equation}
where $ \Delta'$ denotes the derivative of $ \Delta$ with respect to $ r$, and $ r_h$ is a positive real root corresponding to the horizon. In the limiting cases: For $ y \to \infty$, $ Q_m = n = 0$, the metric reduces to the Kerr-Newman metric, which exhibits axial singularities at $ \theta = 0$ and $ \theta = \pi$. For $ y \to \infty$, $ Q_e = Q_m = n = 0$, it reduces to the Kerr metric. For $ y \to \infty$, $ a = Q_m = n = 0$, it becomes the Reissner-Nordström metric. For $ y \to \infty$, $ a = Q_e = Q_m = n = 0$, it simplifies to the Schwarzschild metric.

\section{Charged spin half particle emission from the hot NUT-Kerr-Newman-Kasuya-Anti-de Sitter black hole}\label{sec3}

\begin{figure}
    \centering
    \includegraphics{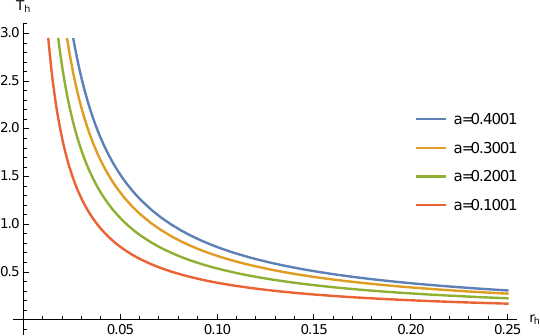}
    \caption{Plot of Hawking temperature near the event horizon of the black hole for the different values of angular momentum at $n =0.674, Q =0.001, y =100.$}
    \label{fig:Haw}
\end{figure}

\begin{figure}
    \centering
    \includegraphics{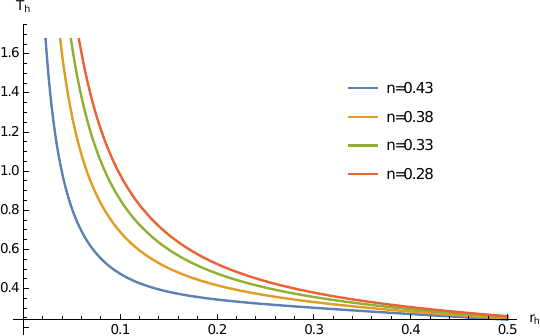}
    \caption{Plot of Hawking temperature near the event horizon of the black hole for the different values of cosmological constant at $n =0.8, Q =0.3, a =97.$}
    \label{fig:magmass}
\end{figure}

\begin{figure}
    \centering
    \includegraphics{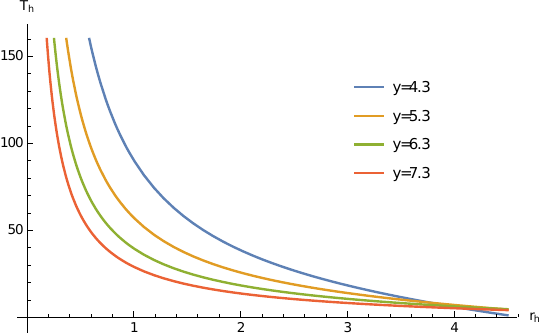}
    \caption{Plot of Hawking temperature near the event horizon of the black hole for the different values of magnetic mass at $a =0.2001, Q =0.1, y =997.$}
    \label{fig:magmass1}
\end{figure}

\begin{figure}
    \centering
    \includegraphics{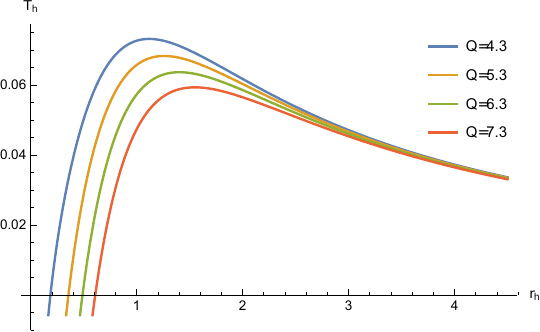}
    \caption{Plot of Hawking temperature near the event horizon of the black hole for the different values of electric and magnetic charges at $a =0.7, y =97, n =0.79.$}
    \label{fig:charge}
\end{figure}

The Dirac equation for a charged spin-$\frac{1}{2}$particle in curved spacetime, as derived in \cite{singha}, is given by:

\begin{equation}\label{dirac}
i\hbar \gamma^{\mu} e^{\nu}_{\mu} D_{\nu} \chi + m \chi = 0, 
\end{equation}
where $ m$ is the mass of the particle, $\chi$ denotes the wave function, and the covariant derivative $ D_{\nu}$ is defined as:
\begin{equation}
\begin{aligned}
D_{\nu} &= \partial_{\nu} + \Omega_{\nu} + \frac{iq}{\hbar} A_{\nu},\\ \quad \Omega_{\nu} &= \frac{1}{2} i \Gamma^{\alpha \beta}_{\nu} \Sigma_{\alpha \beta},\\ \quad 
\Sigma_{\alpha \beta}& = \frac{1}{4} i [\gamma^{\alpha}, \gamma^{\beta}].
\end{aligned}
\end{equation}
Here $ q$ represents the charge of the particle, $ A_{\nu}$ is the electromagnetic potential, $\Omega_{\nu}$ encapsulates the spin connection, $\Gamma^{\alpha \beta}_{\nu}$ are the Christoffel symbols, and $\Sigma_{\alpha \beta}$ are the spin matrices. 
We choose the four tetrad representations in the following form:
\begin{eqnarray}
\begin{aligned}
e^{\nu}_{0}&=\Big(\sqrt{\frac{g_{\phi\phi}}{\Delta}},0,0,\sqrt{\frac{g_{\phi\phi}}{\Delta}}\frac{g_{t\phi}}{g_{\phi\phi}}\Big),\\ \quad
e^{\nu}_{1}&=\Big(0,\frac{1}{\sqrt{g_{rr}}},0,0\Big),\\ \quad
e^{\nu}_{2}&=\Big(0,0,\frac{1}{\sqrt{g_{\theta\theta}}},0\Big),\\ \quad
e^{\nu}_{3}&=\Big(0,0,0,\frac{1}{\sqrt{g_{\phi\phi}}}\Big).\label{tetra}
\end{aligned}
\end{eqnarray}
The chiral $\gamma$ matrices are 
\begin{eqnarray}
	\gamma^{0}=\begin{pmatrix}
		i & 0 \\
		0 & -i
\end{pmatrix},~\gamma^{1}=\begin{pmatrix}
		0 & \sigma^3 \\
		\sigma^3 & 0 
\end{pmatrix},~~\gamma^{2}=\begin{pmatrix}
		0 & \sigma^2 \\
		\sigma^2 & 0 
	\end{pmatrix},\nonumber
	~\gamma^{3}=\begin{pmatrix}
		0 & \sigma^1 \\
		\sigma^1 & 0 
	\end{pmatrix},\nonumber
\end{eqnarray}
and the Pauli matrices $\sigma$'s are
\begin{eqnarray}
	\sigma^{1}=\begin{pmatrix}
		0 &1 \\
		1 & 0 
	\end{pmatrix},~\sigma^{2}=\begin{pmatrix}
		0 & -i \\
		i & 0 
	\end{pmatrix},~\sigma^{3}=\begin{pmatrix}
		1 & 0 \\
		0 & -1 
	\end{pmatrix}.\nonumber
\end{eqnarray}
We write only the spin-up function, $\chi_{\uparrow}$ for the Dirac field eqn. (\ref{dirac}) as
\begin{eqnarray}
\chi_{\uparrow}(t,r,\theta,\phi)=
\begin{bmatrix}
A(t,r,\theta,\phi)\\
0\\
B(t,r,\theta,\phi)\\
0
\end{bmatrix}e^{\frac{i}{\hbar}S(t,r,\theta,\phi)}.\label{wave}
\end{eqnarray}
The expansion of the Dirac field of eqn. (\ref{dirac}) is given by
\begin{eqnarray} 
\begin{aligned}&-i\hbar \bigg (\gamma ^0 e^t_0\partial _t+\gamma ^0 e^\phi _0\partial _\phi +\gamma ^1 e^r_1\partial _r+\gamma ^2 e^\theta _2\partial _\theta +\gamma ^3 e^\phi _3\partial _\phi
 \\&+\gamma ^0 e^t_0\frac{iq}{\hbar }A_t +\gamma ^0 e^\phi _0\frac{iq}{\hbar }A_\phi +\gamma ^3e^\phi _3\frac{iq}{\hbar }A_\phi \bigg )\chi_{\uparrow} =0.
\end{aligned} 
\end{eqnarray}
We employ Dirac's partial differential operator notation, $\partial_{\mu}$. Applying the WKB approximation, we substitute \ref{wave} into \ref{dirac} for the Dirac field. Dividing through by the exponential term and retaining only the leading order terms in $\hbar$, we obtain:
\begin{eqnarray} 
&&A \bigg \{i(e^t_0\partial _t+e^\phi _0\partial _\phi ){\mathcal {S}}+ie^t_0qA_t+ie^\phi _0qA_\phi \bigg \}+B e^r_1\partial _r{\mathcal {S}}=0,\label{eqn1}\\
&&B (ie^\theta _2 \partial _\theta {\mathcal {S}}+e^\phi _3\partial _\phi {\mathcal {S}}+qe^\phi _3A_\phi )=0,\\ &&A e^r_1\partial _r{\mathcal {S}}-B \bigg \{i(e^t_0\partial _t +e^\phi _0\partial _\phi ){\mathcal {S}}+ie^t_0qA_t+ie^\phi _0qA_\phi \bigg \}=0,\label{eqn3}\\ 
&&A (ie^\theta _2\partial _\theta {\mathcal {S}}+e^\phi _3\partial _\phi {\mathcal {S}}+qe^\phi _3A_\phi )=0.  
\end{eqnarray}
Here, $A$ and $B$ are not constants. In applying the WKB approximation, we have omitted all higher-order terms of $O(\hbar)$  in the expansion of the derivatives of $A$ and $B$, as well as the components of $\Omega_{\nu}$. Focusing on the first and third equations, namely \ref{eqn1} and \ref{eqn3}, it is clear that these equations yield a non-trivial solution for $A$ and $B$ only if the determinant of the coefficient matrix vanishes. As a result, we obtain:

\begin{eqnarray}
	\begin{aligned} \bigg (e^t_0\partial _t{\mathcal {S}}+e^\phi _0\partial _\phi {\mathcal {S}}+e^t_0qA_t+e^\phi _0qA_\phi \bigg ) ^2- \bigg (e^r_1\partial _r{\mathcal {S}}\bigg )^2=0. \label{act}
 \end{aligned}
\end{eqnarray}
It is seen that the hot NUT-Kerr-Newman-Kasuya-Anti-de Sitter spacetime possesses two Killing vectors, $(1, 0, 0, 0)$ and $(0, 0, 0, 1)$. We can employ variable separation for the action $\mathcal{S}$ in case of the two decoupling $(m=0)$ equations in the following manner:
\begin{eqnarray}
	\begin{aligned} 
		{\mathcal {S}}(t,r,\theta ,\phi )=-\omega t+J\phi +{\mathcal {R}}(r,\theta ), 
	\end{aligned} \label{action}
\end{eqnarray}
where $\omega$ and $J$ are the Dirac particle's energy and angular momentum. Substituting the given expression for $\mathcal{S}(t,r,\theta,\phi)$ into \ref{act}, we obtain the following result,
\begin{eqnarray}
	\begin{aligned} 
		\bigg (e^t_0\omega-e^\phi _0{J}-e^t_0qA_t-e^\phi _0qA_\phi \bigg)^2-\bigg (e^r_1\partial _r{\mathcal {R}}\bigg )^2=0. 
	\end{aligned}
\end{eqnarray}
Now, we solve the above equation for $\theta=\frac{\pi}{2}$ using the \ref{tetra} and obtain as,
\begin{eqnarray}
	\begin{aligned} 
		\mathcal {R_\pm }&=\pm \int \frac{\Big( e^t_0(\omega-qA_t)-e^\phi _0(J+qA_\phi) \Big )}{e^r_1}dr,\\
		&=\pm \int \frac{\sqrt{\Sigma}}{\Delta\sqrt{g_{\phi\phi}}} \Big(g_{\phi\phi}(\omega-qA_t)-g_{t\phi}(J+qA_\phi) \Big )dr+C, \label{int}
	\end{aligned}
\end{eqnarray}
where C is a complex constant of integration. Here, the $+$ sign represents the outgoing particle away from the black hole, whereas the $-$ sign represents the incoming particle towards the black hole. The integral of \ref{int} has only one pole at the event horizon $r_{h}$. Thus, the imaginary part of $\mathcal {R_\pm }$ is obtained as
\begin{eqnarray}
	\begin{aligned} 
		Im\mathcal {R_\pm }
	&=\mp 2\pi\frac{a^2+r_h^2+n^2}{\Delta'\Xi}\Big(\omega-\omega_{0}\Big)-Im(C)
	\end{aligned} 
\end{eqnarray}
where $\omega_{0}=-\frac{q Q r_h \left(a^2 (\Xi +1)+n^2\right)+a J \Xi ^2 r_h^2+a J n^2 \Xi ^2+q Q r_h^3}{\Xi  \left(r_h^2+n^2\right) \left(a^2+r_h^2+n^2\right)}$.
The imaginary part of the action \ref{action} yields the imaginary part of the radial wave expression. Using the classical limit, the probability, $\Gamma$, of crossing the event horizon in each direction is proportional to 
\begin{eqnarray}
	\begin{aligned} 
	\Gamma_{out}\propto\ e^{Im\mathcal {I_\pm }}=e^{-(ImR_{+}+Im(C))},~~\Gamma_{in}\propto e^{Im\mathcal {I_\pm }}=e^{-(ImR_{-}+Im(C))}.~~
	\end{aligned} \nonumber
\end{eqnarray}
Since any incoming particle that crosses the event horizon has a 100\% chance of entering the black hole, the probability of a particle transitioning from outside to inside the horizon is equal to 1. This holds true only when $ \text{Im}(R_{-}) = -\text{Im}(C)$. Therefore, the probability of a particle tunneling from inside to outside the horizon is given by:
\begin{eqnarray}
	\begin{aligned} \Gamma\propto\frac{\Gamma_{out}}{\Gamma_{in}}&=\frac{e^{-\Big(ImR_{+}+Im(C)\Big)}}{e^{-\Big(ImR_{-}+Im(C)\Big)}}=e^{-\Big(ImR_{+}-ImR_{-}\Big)},\\
		&=e^{-2ImR_{+}}.
	\end{aligned} \nonumber
\end{eqnarray}
It is assumed that the radiation is not purely thermal. It is known while employing the WKB approximation that the tunneling rate is related to the energy and the Hawking temperature of the relative particle as $\Gamma\propto e^{-\frac{\Delta\omega}{T}}$. If $\Delta\omega<0$ is the energy of the emitted particle, then the energy of the outgoing particle must be $-\Delta\omega$ due to the energy conservation. Thus, the probability of tunneling particle across the event horizon will read as $\Gamma\propto e^{\frac{\Delta\omega}{T}}=e^{\frac{\omega-\omega_{0}}{T}}$. Subsequently, the Hawking radiation denoted by $T_h$ is obtained as
\begin{eqnarray}
\begin{aligned} 
T_h&=\frac{\Delta'\Xi}{4\pi(a^2+n^2+r_{h}^2)},\\
&=\frac{\Xi\left(r_h^2 \left(a^2+6 n^2+y^2\right)-a^2 \left(5n^2+y^2\right)+3 r_h^4-5n^4+n^2y^2-Q^2y^2\right)}{4\pi y^2r_h\left(a^2+r_h^2+n^2\right)}. \label{ht}
\end{aligned}
\end{eqnarray}
The Hawking temperature variation near the black hole's event horizon concerning different parameters is shown in \ref{fig:Haw}. It can be observed that the Hawking temperature decreases as the event horizon increases and shifts towards a positive value as the angular momentum increases while keeping the magnetic mass, the black hole's charge, and the cosmological constant fixed. This suggests that the Hawking temperature is influenced by the black hole's parameters, such as angular momentum, magnetic mass, charge, and cosmological constant. \ref{fig:magmass} illustrates how the Hawking radiation varies with the event horizon for different cosmological constant values. \ref{fig:magmass1} shows how the Hawking radiation varies with the event horizon for different values of the magnetic mass. Meanwhile, \ref{fig:charge} presents a typical plot showing how the Hawking temperature changes with the event horizon for various values of the electric and magnetic charges. The Hawking temperature initially increases exponentially, reaching a maximum before decaying, with the behavior depending on the values of the charges. The critical behavior of black holes with a NUT charge is not as well understood as that of Kerr-Newman black holes, but the NUT parameter (n) and the angular momentum (a) influence the spacetime geometry, including the
horizon structure and the complex thermodynamic properties of the black hole, leading to instability depending on the interplay between n, mass M , and angular momentum a of the black hole. The NUT charge modifies the gravitational
potential in such a way that the rate of evaporation differs from a black hole without a NUT charge, potentially affecting the emission of Hawking radiation. This process results in either an increased or decreased evaporation rate depending on the detailed structure of the spacetime, while the NUT parameter n doesn’t have as direct an impact on the evaporation process as angular momentum. The Schwarzschild black hole’s evaporation rate is well-known, with the
Hawking temperature depending solely on its mass. The absence of rotation or charge means simpler dynamics but less flexibility in altering thermodynamic properties. The presence of the NUT parameter could lead to different boundary conditions at infinity, thereby affecting the radiated energy and temperature.\\

\begin{figure}
    \centering
    \includegraphics[width=0.5\linewidth]{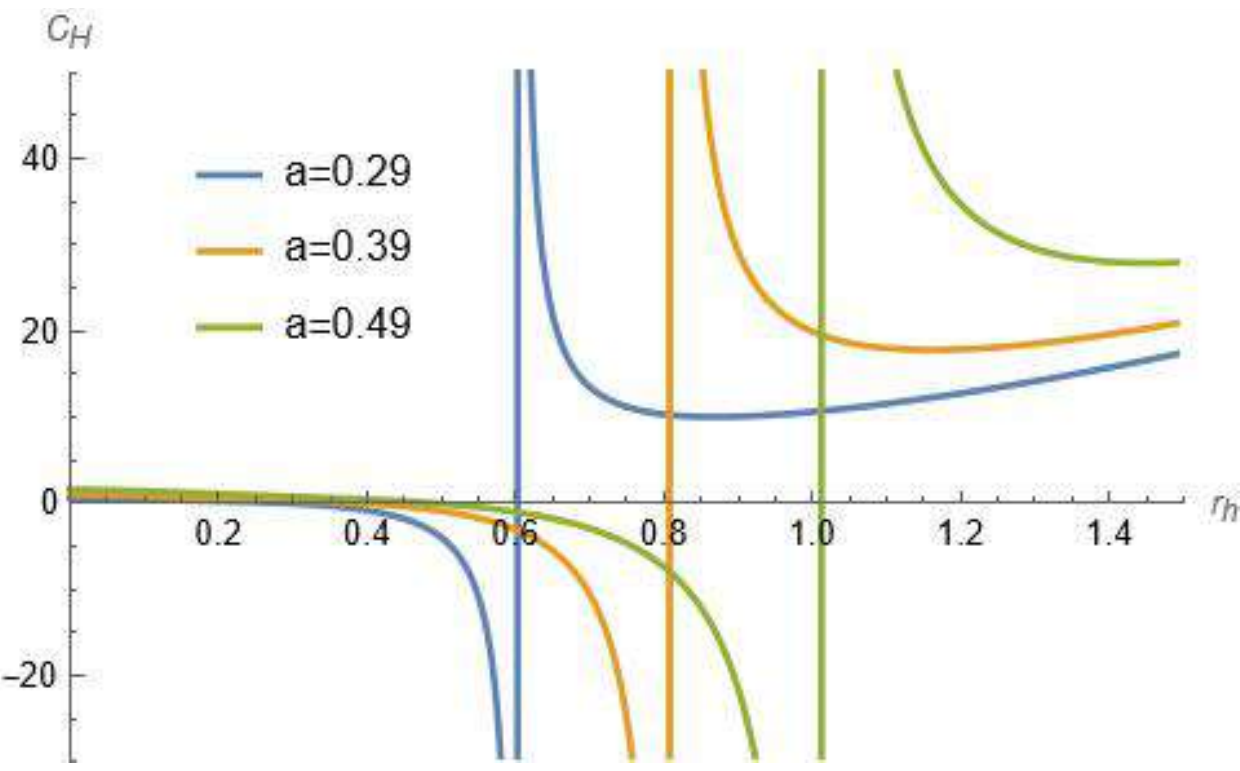}
    \caption{Variation of heat capacity versus the horizon radius for various values of ``a"}
    \label{fig1}
\end{figure}

\begin{figure}
    \centering
    \includegraphics[width=0.5\linewidth]{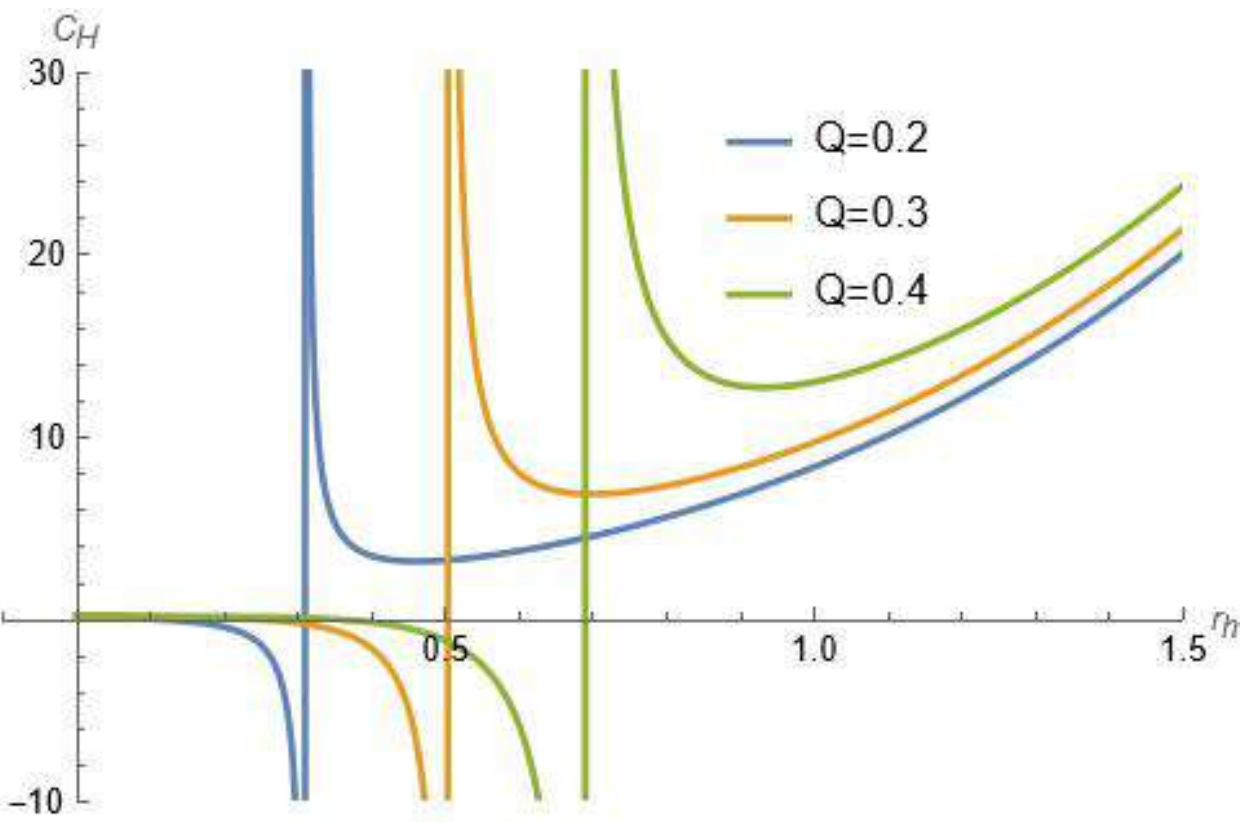}
    \caption{Variation of heat capacity versus the horizon radius for various values of ``Q"}
    \label{fig2}
\end{figure}

\begin{figure}
    \centering
    \includegraphics[width=0.5\linewidth]{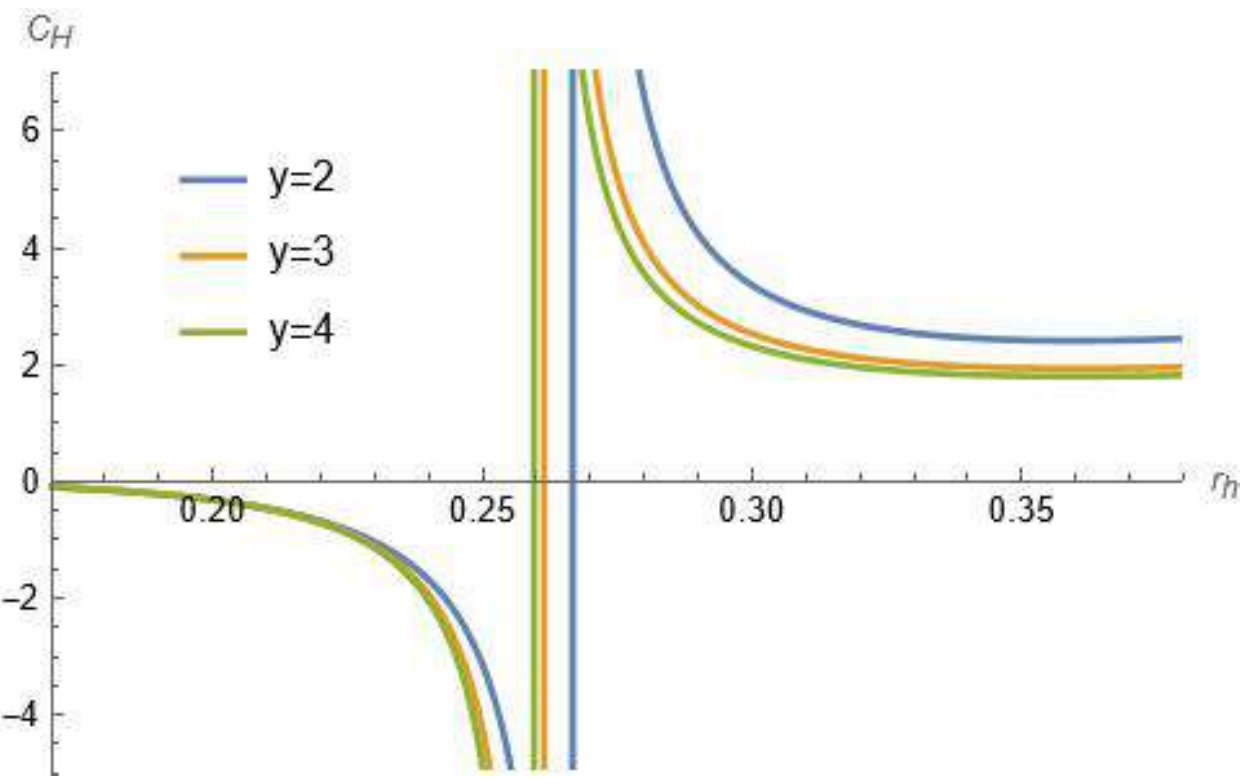}
    \caption{Variation of heat capacity versus the horizon radius for various values of ``y"}
    \label{fig3}
\end{figure}

To get the specific heat, $C_{H}$ of the  HNKNK-AdS black hole, we obtain the black hole mass from $\Delta |_{r=r_h}=0$ as
\begin{eqnarray}
M=\frac{r_h^2 \left(a^2+6 n^2+y^2\right)+a^2 \left(5 n^2+y^2\right)+r_h^4+5 n^4-n^2 y^2+Q^2 y^2}{2 y^2 r_h}.
\end{eqnarray}
Using the mass of the black hole, the heat capacity, $C_{H}$ of the black hole near the event horizon is obtained as
\begin{eqnarray}
\begin{aligned}
C_{H}=\Big(\frac{\partial M}{\partial r_{h}}\Big)\Big(\frac{\partial r_{h}}{\partial T}\Big)=-\frac{2\pi A \left(a^2+r_h^2+n^2\right)^2 }{\Xi B},
\end{aligned}
\end{eqnarray}
where 
\begin{eqnarray}
A&=&-r_h^2\left(a^2+6n^2+y^2\right)+a^2 \left(5 n^2+y^2\right)-3 r_h^4+5 n^4-n^2y^2+Q^2y^2,\nonumber \\ 
B&=&r_h^4 \left(8 a^2+3 n^2-y^2\right)+\left(a^2+n^2\right) \left(5a^2n^2+a^2y^2+5 n^4-n^2y^2+Q^2 y^2\right)\nonumber \\
&&+r_h^2 \left(a^4+22a^2n^2+4a^2y^2+21n^4-2n^2y^2+3 Q^2 y^2\right)+3 r_h^6. \nonumber
\end{eqnarray}
The plot of the heat capacity $C_{H}$ of the black hole near the event horizon is shown in \ref{fig1}. Variations of heat capacity versus the horizon radius for various values of black hole parameters are shown in \ref{fig2}, and \ref{fig3}.
The figures show that the heat capacity is affected by the black hole's parameters: ``a", ``n", ``Q" and the cosmological constant,``y" which are linked to changes in the thermodynamic state of the black hole. Phase transitions occur when there is a sudden change in the black hole parameters, leading to discontinuities.These discontinuities highlight crucial aspects of black hole thermodynamics, indicating transitions between different equilibrium states, much like phase changes in ordinary thermodynamic systems.
It is clearly seen that the heat capacity has infinite points of discontinuity. It is obvious that the smaller the inner event horizon of the black hole, the more the black hole system becomes unstable and unphysical. The black hole system tends to become stable as the event horizon increases.\\
\noindent The Hawking temperature of \ref{ht} for the Hot NUT-Kerr-Newman-Kasuya-Anti-de Sitter Black Hole, which is slightly different from the Kerr-Newman Anti-de Sitter Black Hole due to the additional parameters of the black hole, is corrected 
as
\begin{eqnarray}
    T_h\sim &&\frac{-a^2+r_h^2-Q_e^2-Q_m^2}{4\pi r_h \left(a^2+r_h^2\right)}+\frac{a^4+a^2(Q_e^2 + Q_m^2)+3 r_h^4}{4 \pi  y^2 r_h \left(a^2+r_h^2\right)}\nonumber \\
    &&+n^2\left(\frac{2 a^2+Q_e^2 + Q_m^2}{4\pi r_h\left(a^2+r_h^2\right){}^2}+\frac{-7 a^4-a^2(Q_e^2 + Q_m^2)+3 r_h^4}{4 \pi  y^2 r_h \left(a^2+r_h^2\right){}^2}\right)+...
\end{eqnarray}
The Hawking temperature for the Kerr-Newman black hole 
reduces as $Q_{m}\rightarrow 0, n\rightarrow 0,\text{and}$ $ y\rightarrow \infty$ to
\begin{eqnarray}
   T_{h}=-\frac{a^2+Q_{e}^2-r_h^2}{4\pi r_h(a^2 +r_h^2)},
\end{eqnarray}
which is consistent with the work result. Whereas the heat capacity of the Kerr-Newman black hole is obtained in the limit $Q_{m}\rightarrow 0, n\rightarrow 0,\text{and}$ $ y\rightarrow \infty$ as
\begin{eqnarray}
  C_{H}=-\frac{2\pi(a^2+Q_{e}^2-r_h^2)(a^2+r_h^2)^2}{a^2(a^2+Q_{e})^2+r_{h}^2(4a^2+3Q_{e}^2)-r_{h}^4}.
\end{eqnarray}

\section{Discussion and conclusion}\label{sec6}
In this article, we have derived the Hawking temperature by analyzing the tunneling of a Dirac particle from the black hole. The resulting Hawking temperature is influenced by various parameters of the black hole, including its electric and magnetic charges, magnetic mass, cosmological constant, and the angular momentum of the hot NUT-Kerr-Newman-Kasuya-Anti-de Sitter (HNKNK-AdS) black hole. We also examined the heat capacity near the event horizon, finding that different angular momentum values exhibit infinite discontinuities, which reduce the Hawking temperature for the Kerr-Newman black hole. It is seen that the smaller the inner event horizon of the black hole, the more the black hole system becomes unstable and unphysical. The black hole system tends to become stable as the event horizon increases. Additionally, we calculated the heat capacity of the Kerr-Newman black hole.

\section*{Acknowledgements}

The first two authors acknowledge IUCAA, Pune, for providing local hospitality and computational facilities to carry out this research work. The third author acknowledges the authority of Rajiv Gandhi University to provide the necessary facilities and support to carry out this work.\\


\begin{thebibliography}{10}
	
\bibitem{Hawking1975} Hawking, Stephen W. ``Particle creation by black holes." Communications in mathematical physics 43.3 (1975): 199-220.
	
\bibitem{Gibb} Gibbons, Gary W., and Stephen W. Hawking. ``Cosmological event horizons, thermodynamics, and particle creation." Physical Review D 15.10 (1977): 2738.	
	
\bibitem{Parikh} Parikh, Maulik K., and Frank Wilczek. ``Hawking radiation as tunneling." Physical review letters 85.24 (2000): 5042.
	
\bibitem{Kraus} Kraus, Per, and Frank Wilczek. ``Self-interaction correction to black hole radiance." Nuclear Physics B 433.2 (1995): 403-420.

\bibitem{Angheben} Angheben, Marco, et al.``Hawking radiation as tunneling for extremal and rotating black holes." Journal of High Energy Physics 2005.05 (2005): 014.

\bibitem{Padhmanabhan} Padmanabhan, Thanu. ``Thermodynamical aspects of gravity: new insights." Reports on Progress in Physics 73.4 (2010): 046901.

\bibitem{Christensen}Christensen, Steven M., and Stephen A. Fulling. ``Trace anomalies and the Hawking effect." Physical Review D 15.8 (1977): 2088.

\bibitem{zhang}Zhang, Jingyi, and Zheng Zhao. ``Hawking radiation of charged particles via tunneling from the Reissner-Nordström black hole." Journal of High Energy Physics 2005.10 (2005): 055.
	
\bibitem{zheng}Ma, Zheng Ze. ``Hawking temperature of Kerr–Newman–AdS black hole from tunneling." Physics letters B 666.4 (2008): 376-381.

\bibitem{AliRi} Ali, Riasat, Xia Tiecheng, and Rimsha Babar. ``Study of first-order quantum corrections of thermodynamics to a Dyonic black hole solution surrounded by a perfect fluid." Nuclear Physics B 1008 (2024): 116710.

\bibitem{Sakallı} Sakallı, İzzet, and Esra Yörük. "Modified Hawking radiation of Schwarzschild-like black hole in bumblebee gravity model." Physica Scripta 98.12 (2023): 125307.

\bibitem{Annamalai} Annamalai, D., Pandey, A. ``Tunneling of Hawking Radiation in Starobinsky–Bel–Robinson Gravity". Gravit. Cosmol. 30, 450–454 (2024).

\bibitem{Siahaan} Siahaan, Haryanto M., et al. ``Properties of Melvin–Taub–NUT spacetime with Manko–Ruiz parameter." General Relativity and Gravitation 55.10 (2023): 113.

\bibitem{Liu} Liu, Tin-Ping, Qun-Chao Ding, and Shu-Zheng Yang. ``Study on the Influence of Lorentz Dispersion Relation on Fermions Tunneling Radiation of Kerr-TAUB-NUT Black Hole." International Journal of Theoretical Physics 59 (2020): 3015-3022.
    
\bibitem{Keiju} Murata, Keiju, and Jiro Soda. ``Hawking radiation from rotating black holes and gravitational anomalies." Physical Review D 74.4 (2006): 044018.

\bibitem{wu2000} Wu, S. Q., and X. Cai. ``Generalized laws of black hole thermodynamics and quantum conservation laws on Hawking radiation process." arXiv preprint gr-qc/0004040 (2000).
	
\bibitem{wu2004} Wu, Shuang-Qing. ``The conformal version of black hole thermodynamics." arXiv preprint hep-th/0411002 (2004).
	
\bibitem{wu2005}Wu, Shuang-Qing. ``New formulation of the first law of black hole thermodynamics: a stringy analogy." Physics Letters B 608.3-4 (2005): 251-257.
	
	
\bibitem{paddy1} Padmanabhan, T. ``Essay: The holography of gravity encoded in a relation between entropy, horizon area, and action for gravity." General Relativity and Gravitation 34 (2002): 2029-2035.
	
\bibitem{paddy2}Padmanabhan, Thanu. ``Gravity and the thermodynamics of horizons." Physics Reports 406.2 (2005): 49-125.
		
\bibitem{peltola}Peltola, Ari, and Jarmo Mäkelä. ``RADIATION OF THE INNER HORIZON OF THE REISSNER–NORDSTRÖM BLACK HOLE." International Journal of Modern Physics D 15.06 (2006): 817-843.

\bibitem{Frolov}Frolov, Valery P., and Don N. Page. ``Proof of the generalized second law for quasistationary semiclassical black holes." Physical Review Letters 71.24 (1993): 3902.

\bibitem{Wald}Wald, Robert M. `` ``Nernst theorem'' and black hole thermodynamics." Physical Review D 56.10 (1997): 6467.
	
	
\bibitem{Bekenstein}Bekenstein, Jacob D. ``Black holes and entropy." Physical Review D 7.8 (1973): 2333.
	
\bibitem{Bardeen}Bardeen, James M., Brandon Carter, and Stephen W. Hawking. "The four laws of black hole mechanics.``Communications in mathematical physics 31 (1973): 161-170.
	
	
\bibitem{He}He, Yun, Meng-Sen Ma, and Ren Zhao. ``Entropy of black holes with multiple horizons." Nuclear Physics B 930 (2018): 513-523.
	
\bibitem{Ren} Ren, Jun. ``Thermodynamics Properties of the Inner Horizon of a Kerr-Newman Black Hole." International Journal of Theoretical Physics 48 (2009): 2088-2097.
	
\bibitem{RenJun} Ren, Jun. ``Tunneling Effect of Two Horizons from a Reissner-Nordstrom Black Hole." International Journal of Theoretical Physics 48 (2009): 431-440.
	
\bibitem{Volovik:2021upi} G.~E. Volovik, ``{Effect of the inner horizon on the black hole thermodynamics: Reissner\textendash{}Nordstr\"om black hole and Kerr black hole},'' {\em Mod.Phys. Lett. A}, vol.~36, no.~24, p.~2150177, 2021.
	
\bibitem{Volovik:2021iim}G.~E. Volovik, ``{Macroscopic Quantum Tunneling: From Quantum Vortices to Black Holes and Universe},'' {\em J. Exp. Theor. Phys.}, vol.~135, no.~4, pp.~388--408, 2022.
	
\bibitem{Singha:2021dxe}C.~Singha, ``{Thermodynamics of multi-horizon spacetimes},'' {\em Gen. Rel. Grav.}, vol.~54, no.~4, p.~38, 2022.	
	
\bibitem{Choudhury:2004ph}T.~R. Choudhury and T.~Padmanabhan, ``{Concept of temperature in multi-horizon spacetimes: Analysis of Schwarzschild-de Sitter metric},'' {\em Gen. Rel. Grav.}, vol.~39, pp.~1789--1811, 2007.

\bibitem{Juan}Juan Yang and Shu-Zheng Yang, "Fermions tunneling from rotating stationary Kerr black hole with electric charge and magnetic charge," Journal of Geometry and Physics, vol.~60, pp.~986--990, 2010.

\bibitem{Ali} Ali, M. Hossain, and Kausari Sultana. "Charged Dirac Particles’ Hawking Radiation via Tunneling of Both Horizons and Thermodynamics Properties of Kerr–Newman–Kasuya–Taub–NUT–AdS Black Holes." International Journal of Theoretical Physics 52 (2013): 4537-4556.

\bibitem{Media:2024tyg}
N.~Media, S.~Christina and T.~Ibungochouba Singh, ``Entropy correction of Kerr\textendash{}Newman-AdS black hole in Rastall gravity under Lorentz symmetry violation,'' Int. J. Mod. Phys. A \textbf{39} (2024). 

\bibitem{kasuya} Kasuya, Masahiro. ``Exact solution of a rotating dyon black hole." Physical Review D 25.4 (1982): 995.

\bibitem{singha} C. Singha, et al. "Hawking temperature of black holes with multiple horizons", Gen. Relat. Gravitat. 55(10), 106 (2023)
\end{thebibliography}
\end{document}